\documentclass[twocolumn,showpacs,preprintnumbers,amsmath,amssymb,prl]{revtex4}

\usepackage{graphicx}
\usepackage{dcolumn}
\usepackage{bm}

\begin{document}


\title{
Ferroelectric Soft Phonons, Charge Density Wave Instability 
and Strong Electron-Phonon Coupling in 
BiS$_2$-Layered Superconductors}


\author{T. Yildirim$^{1,2}$}\email{taner@seas.upenn.edu}%
\affiliation{%
$^{1}$NIST Center for Neutron Research, National Institute of Standards and
Technology, Gaithersburg, Maryland 20899, USA
\\$^{2}$Department of Materials Science and Engineering, University of
Pennsylvania, Philadelphia, PA 19104, USA}%

\date{\today}

\begin{abstract} 
 Very recently a new family of layered materials, 
 containing BiS$_2$ planes was 
 discovered to be superconducting  at temperatures up to T$_c$=10 K,
 raising questions about the mechanism of superconductivity in these systems.  
  Here, we present state-of-the-art first principles calculations 
  that directly address this question and reveal several surprising findings. 
  The parent compound LaOBiS$_2$ possesses anharmonic ferroelectric 
  soft phonons at the zone center with a rather large polarization 
  of $\approx 10 \mu C/cm^2$, which is comparable to the well-known 
  ferroelectric BiFeO$_3$. Upon electron doping, new 
  unstable phonon branches appear along the entire line 
  $Q=(q,q,0)$, causing Bi/S atoms to order in a one-dimensional 
  charge density wave (CDW).  We find that BiS$_2$ is a strong 
  electron-phonon coupled superconductor in the vicinity of 
  competing ferroelectric and CDW phases. 
  Our results suggest new directions to tune 
  the balance between these phases and increase 
  T$_c$ in this new class of materials.
\end{abstract}

\maketitle

Superconductivity -- a phenomenon first documented in 1911 -- remains 
one of the most challenging subjects of condensed matter physics. 
Examples of layered superconductors include cuprates\cite{1}, 
MgB$_2$\cite{2}, CaC$_6$\cite{3}, and recent iron-pnictides\cite{4}. 
Very recently a new family of layered materials, Bi$_4$O$_4$S$_3$\cite{5} 
  and RO$_x$F$_{1-x}$BiS$_2$  (R=La, Nd, Pr, and Ce)\cite{6,7,8,9},
containing BiS$_2$  planes was discovered to be superconducting
 at temperatures up to 10 K. These new systems are structurally similar 
 to the layered, iron-based superconductors LaO$_x$F$_{1-x}$FeAs\cite{4}, 
 and in both cases the superconductivity is achieved by F-doping.  
 In addition,  band structure calculations\cite{11,12} indicate the presence of strong 
 Fermi surface nesting at the wave vector $(\pi,\pi)$, which is the 
 hallmark property of the Fe-pnictides\cite{4}.
 These similarities have 
 raised the exciting question of whether or not the superconducting
  mechanism in the BiS$_2$ system is related to that in the iron 
  pnictides and have therefore generated enormous
  interest\cite{5,6,7,8,9,10,11,12,13,14,15,16,17,18}.

The fundamental question is whether or not the observed 
T$_c$ in this new system can be understood within 
a conventional electron-phonon coupling framework, 
or is a more exotic mechanism responsible for the 
superconducting pairing?  In this work, we present 
state-of-the-art first principles calculations that 
directly address this question and reveal several 
surprising findings. 
We show that the parent compound
LaOBiS$_2$ possesses highly anharmonic, ferroelectric, 
soft phonons at the zone center and a spontaneous 
polarization of $\sim 10 \mu C/cm^2$ 
that is comparable to that in the well-known multiferroic 
BiFeO$_3$. 
Upon electron doping new instabilities appear 
along the entire line $Q=(q,q,0)$, which causes the Bi/S 
atoms to order into a one-dimensional charge 
density wave (CDW).  We find that BiS$_2$ is a 
strong electron-phonon coupled superconductor 
in close proximity to competing ferroelectric 
and CDW phases. Our results suggest new 
directions with which to tune the balance 
between these phases and increase
T$_c$ in this new class of materials.

\begin{figure}
\includegraphics[width=6.0cm]{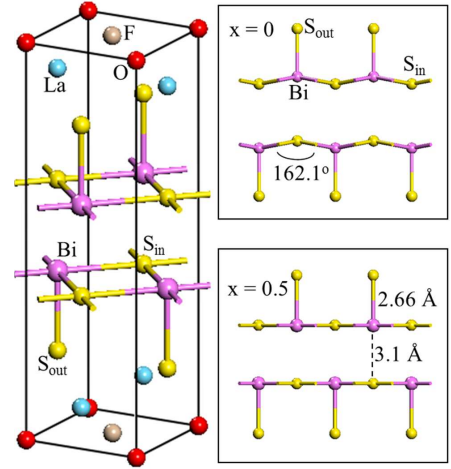} 
\caption{
(color online)
 Crystal structure of LaO$_x$F$_{1-x}$BiS$_2$ (x=0.5). 
 The side views of the BiS$_2$ bilayer for x=0 (top) 
 and x=0.5 (bottom) are also shown. 
 Note the zigzag pattern of the BiS$_2$ plane for x=0.
}
\label{fig1}
\end{figure}
  
Figure~1 shows the P4/nmm tetragonal cell of the LaO$_x$F$_{1-x}$BiS$_2$ which 
consists of two types of atomic layers; namely the LaO spacer and 
electronically active BiS$_2$ bilayer. Upon F-doping in the LaO layer, 
one can control the charge transfer to BiS$_2$ bilayer and thus tune 
the electronic properties. In order to simulate such doping in our 
calculations, we generated $2\times2\times1$ supercell of 
LaOBiS$_2$ and replaced 
some of the oxygen atoms with F atoms in an ordered fashion with 
doping levels of $x=0,0.125,0.25,0.375$, and $0.5$. 
The results are summarized in Fig.~2.

\begin{figure}
\includegraphics[width=7.5cm]{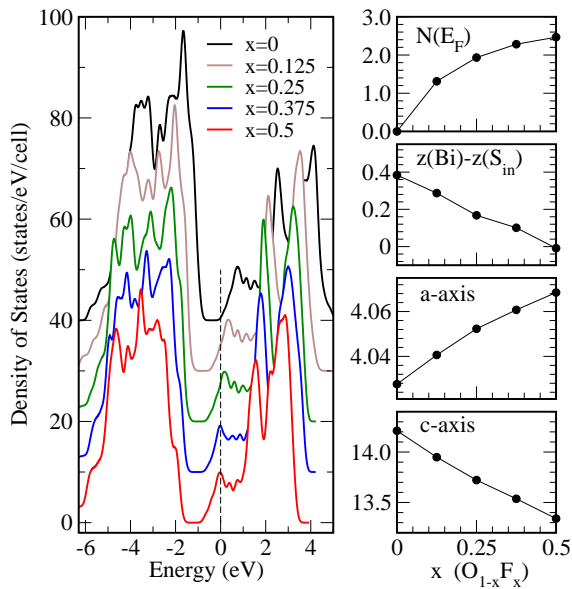} 
\caption{
(color online)
Electronic DOS of LaO$_x$F$_{1-x}$BiS$_2$ as a function of 
F-doping $x$. For clarity the curves are shifted vertically except 
$x=0.5$. On the right, N(E$_F$) (scaled per 10-atom unit cell), 
lattice parameters and the difference of the z-values (all in \AA)
of in-plane S (S$_{in}$) and Bi atoms are also given.
}
\label{fig2}
\end{figure}

The undoped parent compound is a band-insulator with a gap of $\approx 0.8$ eV. 
Upon electron doping, we start to fill the empty states in a rigid-band 
fashion, turning the insulating parent compound into a metallic system. 
From the projected density of states (DOS), we determine that the 
states near the Fermi level are mainly Bi$-6p$ and S$-3p$ character. 
The density of states at the Fermi level, N(E$_F$), increases with 
increasing electron doping and becomes maximal at the half filing x=0.5. 
In agreement with experiments, the c-axis length decreases while 
the a- and b-axis lengths increase with F-doping. Interestingly 
the z-values of the in-plane S (i.e., S$_{in}$) and Bi atoms get closer 
to each other with doping, yielding a nearly perfect planer 
structure at x=0.5. On the other hand zigzag buckled planes 
are formed at x=0, as shown in Fig. 1. This rearrangement of 
Bi and S atoms into a perfect planer structure should have 
important consequences for the electronic band structure and 
the nature of Fermi surface. Hence it is important to keep an 
eye on the degree of buckling of BiS$_2$ plane and its relation 
to T$_c$ as new isostructural materials are being discovered. 
We note that similar buckling of CuO$_2$ plane in cuprates 
with doping were found to be closely correlated with
 superconducting temperature\cite{19}.
 
 \begin{figure}
\includegraphics[width=7.5cm]{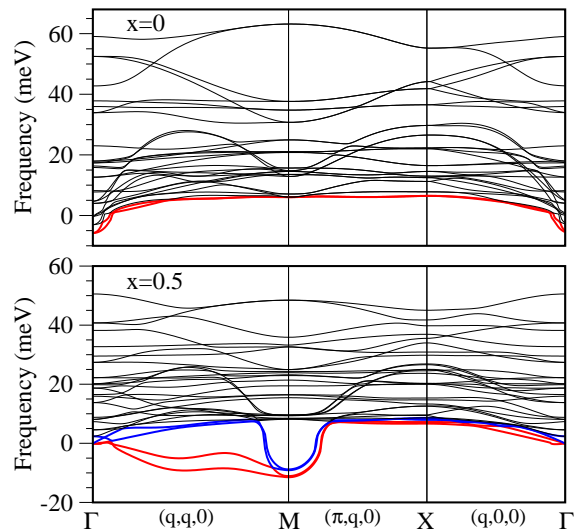} 
\caption{
(color online)
Phonon dispersion curves of LaOBiS$_2$ (top) 
and LaO$_0.5$F$_0.5$BiS$_2$ (bottom), indicating the instabilities 
at $\Gamma$ for $x=0$ and along the entire line $(q,q,0)$
with the most unstable phonons at M and $(\pi/2,\pi/2)$ for 
$x=0.5$.
}
\label{fig3}
\end{figure}
 
One of the most interesting results shown in Fig.~2 is that N(E$_F$) 
is quite high at half-filling (x=0.5) and E$_F$ coincides with a
 peak in the density of states. This usually suggests some 
 sort of instability due to a Van Hove singularity. 
 In order to check this, we calculated phonon dispersion curves 
 of the parent (x=0) and half doped (x=0.5) systems using 
 $4\times4\times1$ supercell and the results are shown in Figure 3. 
 To our surprise, we find that the undoped system shows 
 instability near $\Gamma$ while the doped system has instabilities 
 along the entire line of Q=(q,q,0) (see also Fig. S2).  
 We repeated the phonon dispersion calculations for x=0.5 
 using a charged system without F-doping and obtained the 
 same instabilities (see Fig. S1), indicating that the soft 
 phonons found here are an intrinsic property of the BiS$_2$ plane.

\begin{figure}
\includegraphics[width=7.5cm]{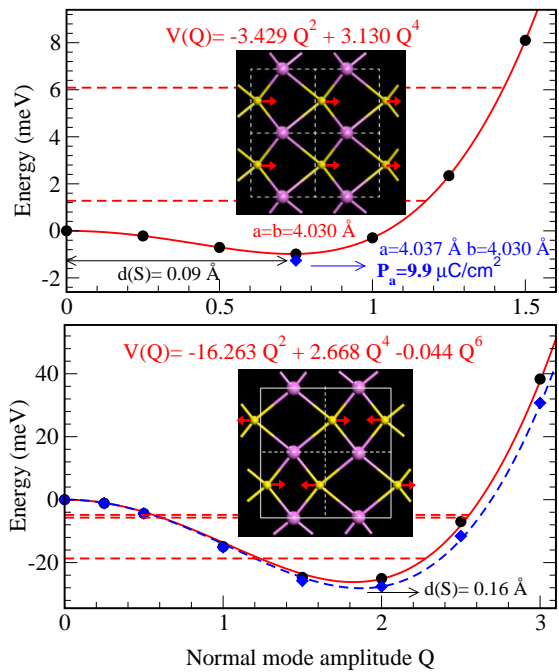} 
\caption{
(color online)
Total energy as the system is distorted by  
the most negative energy phonons in the
 tetragonal cell of LaOBiS$_2$ at $\Gamma$  (top) 
 and in $\sqrt{2}\times\sqrt{2}$ cell of LaO$_{0.5}$F$_{0.5}$BiS$_2$   
  at M$(\pi,\pi)$ (bottom). The insets show the 
  sketch of the unstable phonons. 
  Horizontal dashed red-lines show the energy levels 
  of the frozen-phonon potential (red curve). 
  The calculated polarization, 
lattice parameters and the displacement of S atom are also indicated.
}
\label{fig4}
\end{figure}

In order to have a better insight into the nature of these unstable 
phonons, we carried out total energy calculations as the system 
was distorted by the modes having the most negative energy. 
Figure 4a shows the most unstable soft phonon $E_u$ at $\Gamma$ 
for x=0, which lowers the symmetry from P4/nmm to P21mn (see Fig. S3). 
 The S atoms move towards Bi atoms along a-axis (or b-axis) and slightly 
 lower the energy of the system by $\approx 1$ meV. When the lattice parameters 
 are relaxed, the tetragonal cell becomes orthorhombic and the system 
 energy is further lowered by 0.3 meV. The final structure is shown 
 in Fig. S3. It is remarkable that in the distorted phase, 
 the inversion symmetry is broken and a large spontaneous 
 polarization of $P=9.9 \mu C/cm^2$ is induced despite the rather
  small displacements. The calculated polarization is 
  comparable to that of the well-known room temperature 
  ferroelectric BiFeO$_3$ system\cite{20}. However, we note that 
  the potential curve for this unstable ferroelectric 
  phonon mode is quite shallow, as shown in Figure~4, 
  and quantum zero-point motions may not allow such structural 
  distortion. In fact, solving the Schrödinger equation
   for this potential, we obtained energy levels which are 
   above the potential minimum. Hence, the system should be 
   dynamically disordered due to zero-point motions and 
   should appear as tetragonal. These findings seem to be 
   consistent with the room temperature x-ray data reported so far.  
   Finally, we suggest that the origin of the anharmonic ferroelectric
    mode could be due to mismatch between the optimum lattice 
    parameters of the LaO and the BiS$_2$ layers. If the Bi-S bond 
    is forced to elongate due to interaction between LaO and BiS2 
    planes, then it is natural for the S atom to break the symmetry 
    and move towards the Bi atom to optimize Bi-S interaction. 
    In fact, repeating the phonon calculations for smaller lattice 
    parameters (i.e. a=b=3.8 Å, corresponding to 100 kbar pressure), 
    we do not get negative phonon energies at the zone center any more. 
    Hence, it seems that by changing the spacer oxide, one may tune the
     nature of the soft phonons and the ground state structure.

 \begin{figure}
\includegraphics[width=7.5cm]{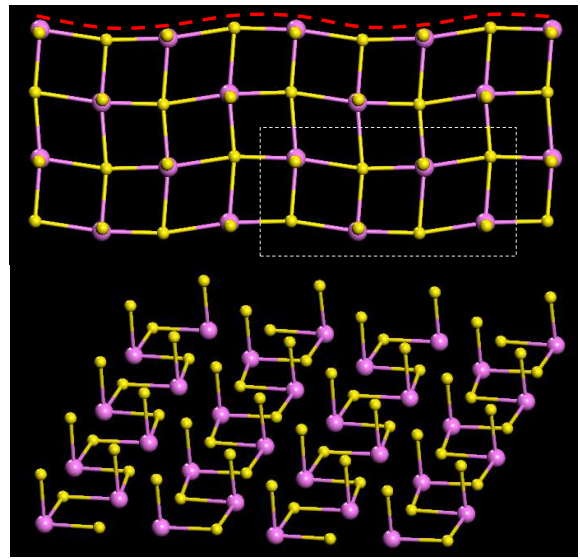} 
\caption{
(color online)
1D Charge Denity Wave formation. The BiS$_2$ plane in fully optimized 
CDW phase of LaO$_{0.5}$F$_{0.5}$BiS$_2$.  Large pink and small yellow 
spheres are Bi and S, respectively. Dashed red line is to guide the eye, 
indicating the sinusoidal distortion of the atoms. 
The white square indicates the $2\sqrt{2}\times\sqrt{2}$ unit cell of 
the CDW phase. Bottom panel shows the same structure with a 
Bi-S bond cutoff distance of 2.8 \AA. The one dimensional nature 
of the chains becomes apparent. The intra-chain Bi-S bond is about 
2.75 \AA~ while inter-chain Bi-S bond is around 3  \AA~  (see Fig. S3).
}
\label{fig5}
\end{figure}

Figure 4b shows the total energy in the $\sqrt{2}\times\sqrt{2}$ cell of the 
tetragonal structure for $x=0.5$ as it is distorted by the 
most negative energy phonon mode at $(\pi,\pi)$. 
Unlike the parent compound,
 the distortion lowers the system energy significantly, causing 
 S atoms move away by 0.16 \AA~ from the high symmetry site. 
 Solving the Schrodinger equation for the resulting potential 
 curve numerically, we obtained the energy levels which are 
 bound to the local minimum of the distortion. The blue-dashed 
 line in Fig. 4b shows that the same distortion also occurs for
  the charged system without F-doping. Hence, the observed 
  soft-mode is an intrinsic property of the BiS$_2$ plane. 
We next checked whether the $(\pi,\pi)$ phonon optimized 
$\sqrt{2}\times\sqrt{2}$ structure 
is stable at other q-vectors by repeating the phonon calculations 
using a $2\sqrt{2}\times2\sqrt{2}$ super-cell. 
We obtained more negative energy 
modes that correspond to the original instability at the $(\pi/2,\pi/2)$
 of the $4\times4\times1$ tetragonal supercell calculations. 
 Hence, we further 
 distorted the  $2\sqrt{2}\times2\sqrt{2}$ structure by the 
 $(\pi/2,\pi/2)$  negative 
 energy phonon and let the system relax. We determined that the 
 optimized structure has the $\sqrt{2}\times2\sqrt{2}$ unit cell with a rather 
 interesting rearrangement of Bi and S atoms in the BiS$_2$ plane,
  as shown in Fig.~5 and Fig.~S4. We call this distorted structure 
  as CDW phase, due to the sinusoidal distortion of the Bi and S 
  atoms as shown by red dashed lines in Figure~5. Unlike the $x=0$ case,
   applying pressure or using smaller lattice constants in the phonon 
   calculations does not stabilize the negative energy phonons.  
   Hence, it is tempting to conclude that the origin of this 
   distortion is due to strong Fermi surface nesting at M\cite{11,12}. 
   However, we note that the soft phonon branch occurs along the 
   entire line $Q=(q,q,0)$ and not just at the M point. Hence,
    in addition to Fermi surface nesting, there should be an 
    important structural reason as well for the observed soft modes. 
    Finally, by looking at the bond distances between Bi-S, we 
    notice that there are one-dimensional channels of Bi-S bonding
     as shown in Fig.~5.  Interestingly, from Hall Effect measurements14 
     it was concluded that superconducting pairing occurs in 
     one-dimensional chains in these systems, which could be 
     related to the CDW phase predicted here. Similarly, high
      pressure measurements reveal non-monotonic dependence of T$_c$ 
      on pressure and suggest that the Fermi surface 
      of LaO$_x$F$_{1-x}$BiS$_2$ 
      could be in the vicinity of instabilities\cite{10}. 
      Finally, the x-ray
       powder diffraction\cite{6} for LaO$_0.5$F$_0.5$BiS$_2$ 
       shows rather unusual
        broad peaks, which could be again related to the CDW phase.

 We next address the nature of superconductivity found in these BiS$_2$ 
 layered systems. We carried out el-ph coupling calculations using
  both $(\pi,\pi)$ phonon optimized $\sqrt{2}\times\sqrt{2}$  
  and CDW structures. We   consider $2\sqrt{2}\times2\sqrt{2}$ supercells, 
  containing 80 atoms and obtain 
  the el-ph coupling by the frozen-phonon method. The results are
   summarized in Figure~6.  In both  $(\pi,\pi)$-optimized or fully distorted 
   CDW structures, the Eliashberg functions are quite similar, 
   indicating that the main physics of the electron-phonon coupling
    mechanism does not depend on the details of the structure.  
    In the $(\pi,\pi)$-phonon optimized structure, N(E$_F$) is higher than 
    in  the CDW phase, and therefore leads to higher el-ph coupling. 
    The total electron-phonon coupling is $\lambda =0.83$ with a logarithmic 
    frequency average $\omega_{log} = 101 $K. These values give a maximum 
    superconducting temperature of T$_c$ = 8.5 K. Even in the fully 
    distorted CDW phase, we get quite large electron-phonon coupling 
    $\lambda =0.6$ and a logarithmic frequency average $\omega_{log} = 122$ K, 
    which gives  a T$_c$ of 6 K. These values are in excellent agreement with the 
    reported experimental values of T$_c$ which vary from 3K to 10 K, 
    depending on the level of doping and sample quality.  
     Inspecting the modes which give the highest el-ph coupling, 
     we estimate that about 90\% of $\lambda$ comes from in-plane Bi and 
     S phonons while the remaining 10\% is due to phonons along the c-axis. 
     There are two bands of phonons near (5 to 10) meV and (15 to 25) meV. 
     The phonons in the lower energy band are due to coupled Bi and 
     S motion while the phonons in the high energy bands are due to 
     almost pure S oscillations. Animation of these modes can be found
      in Ref.24. In Figure~6, we also show the total phonon density
       of states along with the atomic-projections of the DOS. 
       As expected based on the masses of atoms, O and F-based 
       phonons are above 30 meV and do not producing much el-ph 
       coupling. The La phonons are near 10 meV but, as expected, 
       do not produce any significant el-ph coupling.

 \begin{figure}
\includegraphics[width=7.5cm]{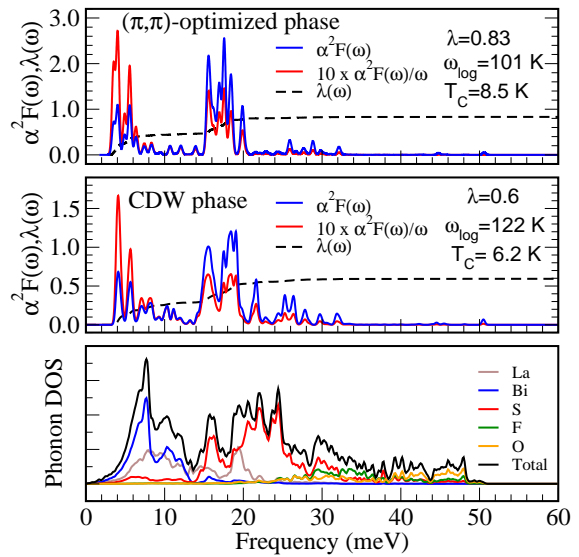} 
\caption{
(color online)
Eliashberg functions in partially optimized (top) and fully optimized 
CDW structure (middle). El-ph coupling constant 
$\lambda, T_c, \omega_{log}$ are also given. 
The total and atomic-projected phonon DOS are 
shown in the bottom panel.  
}
\label{fig6}
\end{figure}

In conclusion, we have discovered rather unusual structural and dynamical 
properties of LaOBiS$_2$ with electron doping. The large-amplitude in-plane 
S-atom displacement controls the structural properties and gives rise 
to large el-ph coupling. It would be interesting to measure the isotope 
effect for the S atom, which may be unconventional. Our results also 
suggest that thin films of BiS$_2$ on various substrates may exhibit 
unusual properties due to epitaxial strain at the interface. 
New materials with similar structures but with BiO$_2$ plane could be 
quite interesting due to smaller mass of oxygen atom, which gives 
higher phonon energies, and in turn higher T$_c$. In fact, the results 
reported here remind us of another interesting system, namely 
Ba$_{1-x}$K$_x$BiO$_3$\cite{25}, which exhibits several complicated structural 
phases\cite{26} and superconductivity at 31~K\cite{25}. It is too early to say 
which of these predictions from DFT calculations will be demonstrated 
experimentally. However it is clear that the BiS$_2$ based layered systems 
are very rich in physics, involving nearly ferroelectric soft phonons 
and CDW ordering along with strongly coupled electron-phonon
superconductivity.

{\it Acknowledgments:}
We acknowledge many fruitful discussions with A. B. Harris and P. M. Gehring.  
TY acknowledges partial support by the U. S. Department of Energy through BES Grant 
No. DE-FG02-08ER46522 for the computational resources used in this study.

{\it Supporting Information:}
The details of methods and additional plots of the phonon dispersion curves, 
picture of optimized structures and atomic positions/lattice parameters are given 
in supporting information (SI).

\setcounter{figure}{0}
\makeatletter 
\renewcommand{\thefigure}{S\@arabic\c@figure} 

\section{Supporting Infomation}

{\bf 1. Methods} The first-principles calculations were performed within the plane-wave implementation 
of the Perdew-Burke-Ernzerhof generalized gradient approximation to density 
functional theory as implemented in the PWSCF package[21].  We used Vanderbilt-type ultrasoft
 potentials. The wavefunction and charge cutoffs are taken as 40 Ry and 480 Ry, respectively. 
 We tested that k-sampling converges well with a k-point grid of 20x20x6 in the 
 tetragonal cell. For $\sqrt{2}\times\sqrt{2} $  structure, we used k-mesh of 14x14x6. 
 The structures are optimized until the forces on atoms are less than 0.005 eV/\AA~  
 and the stress is less than 0.01 kbar.   Some of the total energy and phonon 
 calculations were also repeated by Vienna Ab-Initio Simulation Package 
 using more accurate Projector Augmented Waves method[22]. 
 We obtained very similar results concerning the structural distortion, 
 soft phonons and their energies.   Finally, the electric polarization 
 is calculated by PWSCF via the modern theory of polarization (i.e. Berry Phase).  
 We used 20x20x6 k-mesh for self-consistent calculations and the polarization is 
 calculated using twice dense k-grid along the calculated polarization direction.
The phonon dispersion curves were obtained by the direct finite displacement method. 
We used 0.02 \AA~ displacements in plus and minus directions to obtain the 
force-constant matrix numerically. Once we have the phonon spectrum, 
the electron-phonon coupling is calculated by the supercell frozen-phonon method[23]. 
Briefly, for a given phonon branch, the system is distorted according to the normal 
mode by its root-mean-square (rms) value and then the electron-phonon matrix element, 
$M_{n,n'}  = <n| H_p - H_0| n'>$ , is calculated self-consistently using the wave-functions 
($|n>$) and eigenvalues ($E_n$) of the total potential $H_p$, $H_0$ (i.e., frozen-phonon and 
reference cell, respectively). The el-ph coupling constant $\lambda$ is then obtained from the 
Fermi surface averaging of $M_{n,n'}$ using gaussian smearing method. 
The superconducting critical temperature T$_c$, was estimated using the 
Allen-Dynes formula,  with $\mu^*=0 $ for the upper limit  of  T$_c$.

{\bf 2. Phonon Dispersion Curves in the Electron-Doped System without F-substitution}

\begin{figure}
\includegraphics[width=7.0cm]{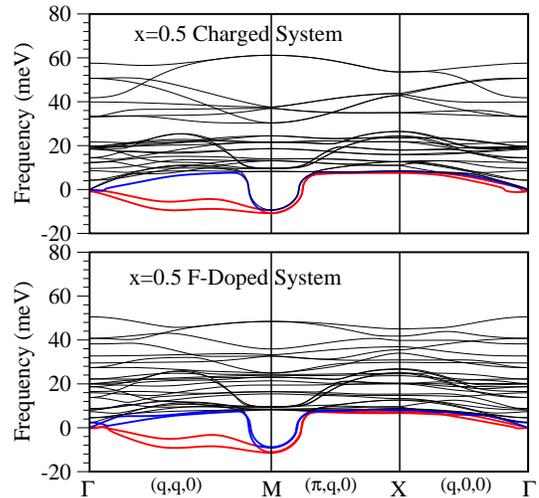} 
\caption{    
Phonon dispersion curves for x=0.5 with charged system without F-doping (top) 
and with F-doped case (bottom).
}
\label{FigS1}
\end{figure}

Here we compare the phonon dispersion curves obtained from $4\times4\times1$ supercell of the 
tetragonal LaOBiS$_2$ cell with $x=0.5$ doping level (top panel in Fig.~S1) 
with the one obtained from actual F-substituted system (bottom panel in Fig.S1). 
Note that both systems show the same phonon instability along the entire line $Q=(q,q,0)$. 
Hence the predicted soft modes do not depend on the details of the doping and it is 
intrinsic to the BiS$_2$ plane. The main difference between the actual 
F-doping and just adding an electron to the system is the maximum phonon energies. 
In the case of charge doping, the top phonon bands extend up to 60 meV 
(which is similar to undoped case). The upper phonon band is somewhat 
separated from the bottom phonon bands. In the case of F-doping, 
the top-phonon band gets softer and mixed up with the lower energy phonon bands. 
This is expected as the F-atom has smaller diameter than oxygen atom and when 
F is substituted at the oxygen site, there will be more room for the surrounding 
atoms to oscillate easily, thus yielding lower phonon energies.

{\bf 3. The soft-phonon branches along the entire $(q,q,0)$ direction for LaO$_x$F$_{1-x}$BiS$_2$}

In Fig.~S2  we present the phonon dispersion curves along $(q,q,0)$ direction which is obtained 
from different supercells, namely $2\times 2\times 1$, 
$3\times 3\times 1$, and $4\times 4\times 1 $ of the LaO$_x$F$_{1-x}$BiS$_2$.  
We also calculated the same dispersion curve using the linear response theory, 
which does not assume any supercell. We note that in $2\times2\times1$  supercell, 
the calculations have the exact phonon energies only at (0,0,0) and $(\pi,\pi,0)$  
and therefore does not reflect the correct dispersion curve.  
For $3\times3\times1$ supercell, we have exact phonon energies at 
$Q=(n 2\pi/3,m 2\pi/3,0) (n,m=0,1,2)$. For $4\times4\times1 $  supercell, 
the exact phonon energies are at $Q=(n2\pi/4,m2\pi/4) (n,m=0,1,2,3)$. In supercell 
calculations the smooth dispersion curves are obtained by assuming a 
real-space cutoff distance for the force-constant matrix. From Figure S2,
 it is clear that we need at least $3\times3\times1$  supercell to get the right dispersion curves. 
 The linear response theory, in excellent agreement with the supercell calculations, 
 also shows that there are two unstable phonon branches along the entire $Q=(q,q,0)$ direction. 
 In Figure S3, we show only the raw-data from linear response theory calculations, 
 without making any approximation for the real-space force constants and cutoff, 
 that is needed to get the smooth curves.

\begin{figure}
\includegraphics[width=7.0cm]{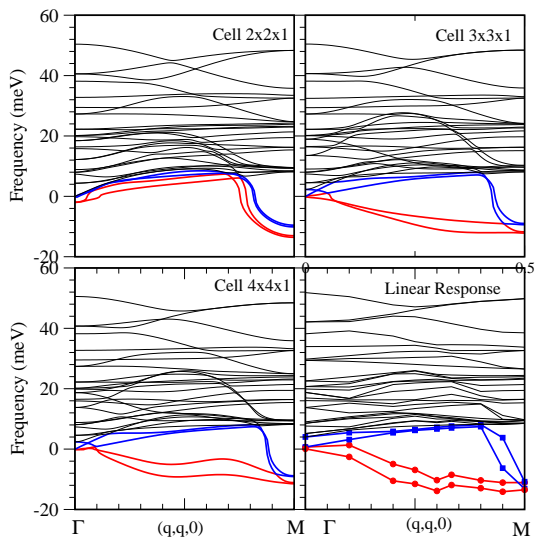} 
\caption{  
Phonon dispersion curves from different supercells and the 
linear response theory, all showing that we have two unstable 
phonon branches along the entire line (q,q,0).
}
\label{FigS2}
\end{figure}

{\bf 4. The Ferroelectric Phase of LaOBiS$_2$ }

Figure S3 shows the structural parameters  
 of the ferroelectric phase of LaOBiS$_2$ after the tetragonal 
 structure is distorted by the most negative energy mode and then fully relaxed. 
 The lattice parameter along the S-displacement direction gets longer, 
 causing one-dimensional Bi-S zigzag chains perpendicular to the 
 S-displacement with intra-chain Bi-S bond distance of 2.8 \AA~ and 
 inter-chain Bi-S bond distance of 2.96 \AA.
   
\begin{figure}
\includegraphics[height=4.0cm]{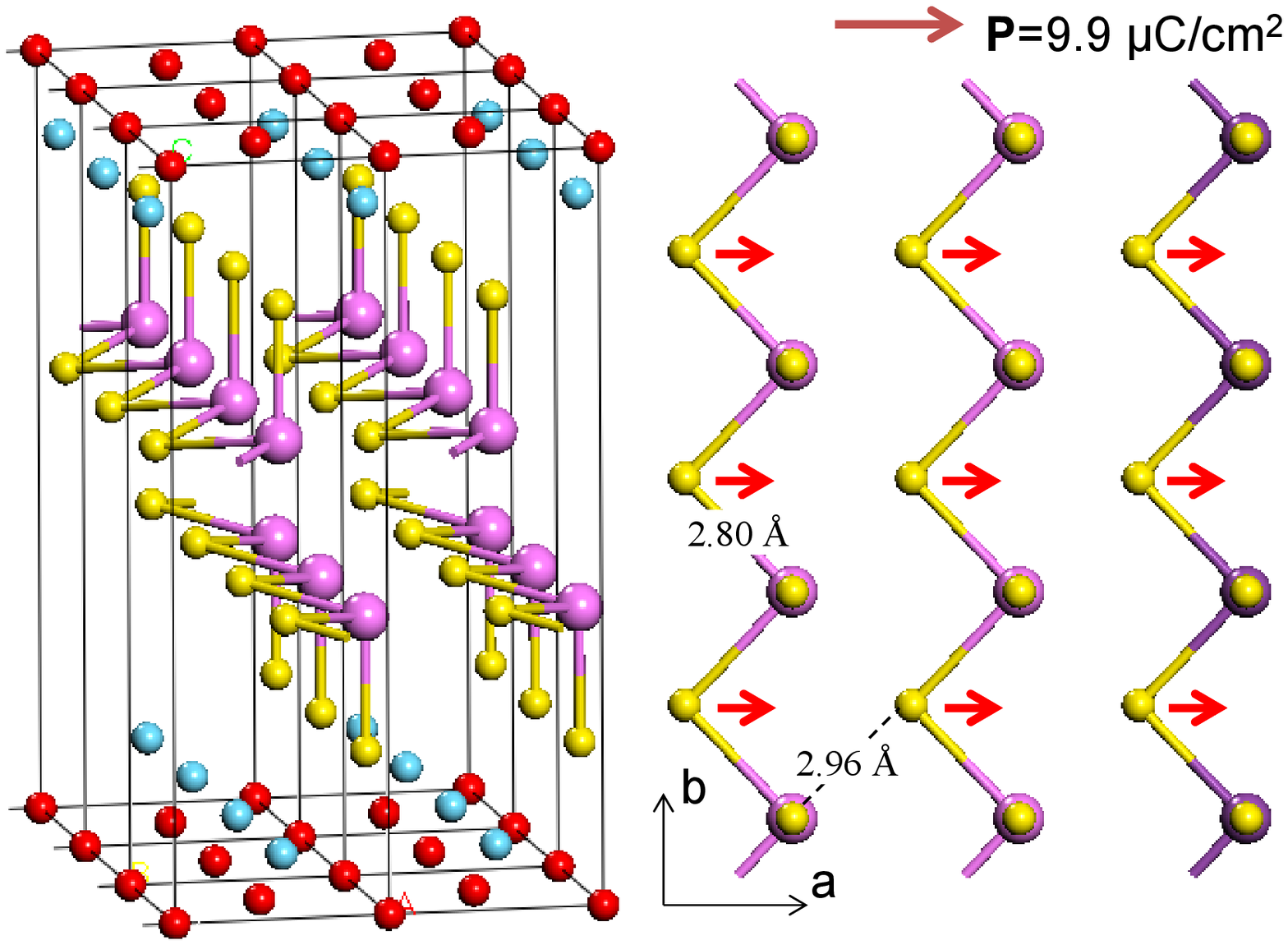} 
\includegraphics[height=4.0cm]{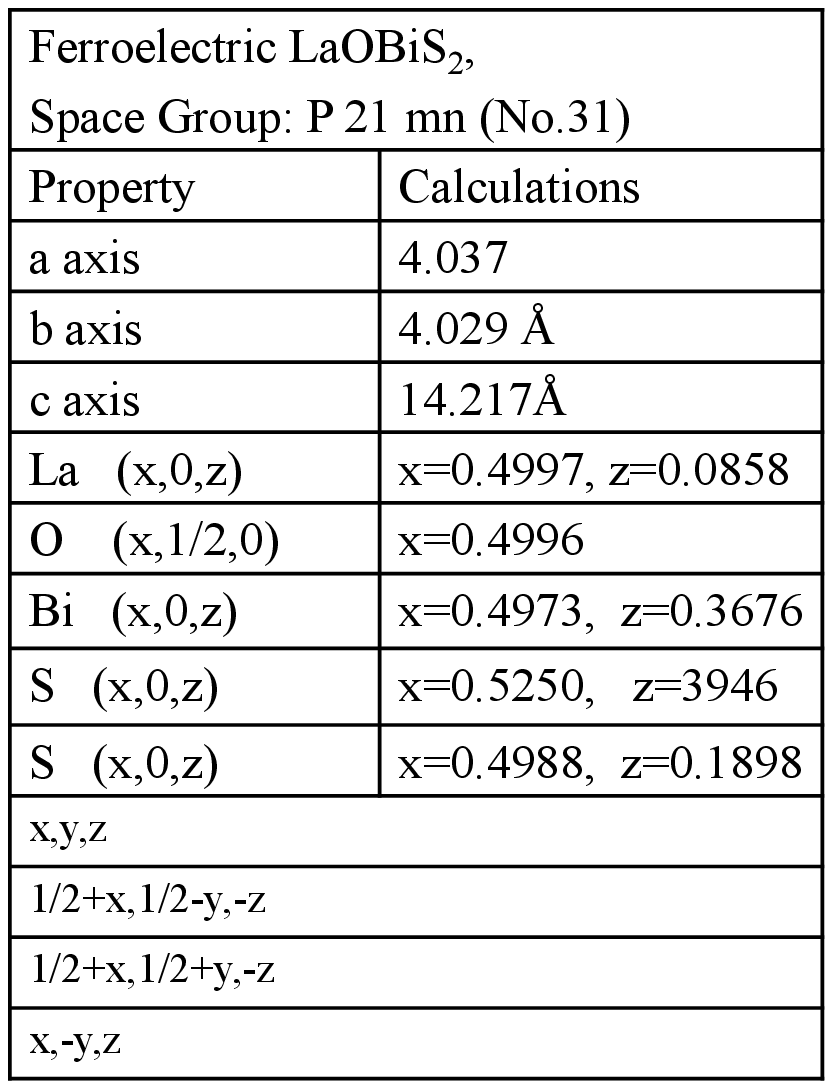}
\caption{
The ferroelectric phase of LaOBiS$_2$ which is obtained by first 
distorting the tetragonal phase using the most unstable gamma-phonon
 and then by performing full structural relaxation.  The bond-stick 
 model shows only the bonds with bond-distance of 2.8 \AA~ to emphasize 
 the one-dimensional chain perpendicular to the in-plane S-displacements 
 (shown by red arrows). The Table on the right shows the optimized 
 structural parameters as well as the symmetry operations of the
  ferroelectric phase.
}
\label{FigS3}
\end{figure}

{\bf 5. The CDW Phase of LaO$_{0.5}$F$_{0.5}$BiS$_2$} 

Figure S4 shows the structural parameters of the CDW phase 
of LaO$_{0.5}$F$_{0.5}$BiS$_2$ after the tetragonal structure is distorted 
by the most negative energy modes at $(\pi,\pi)$ and $(\pi/2,\pi/2)$, 
respectively, and then fully relaxed.

\begin{figure}
\includegraphics[height=5.0cm]{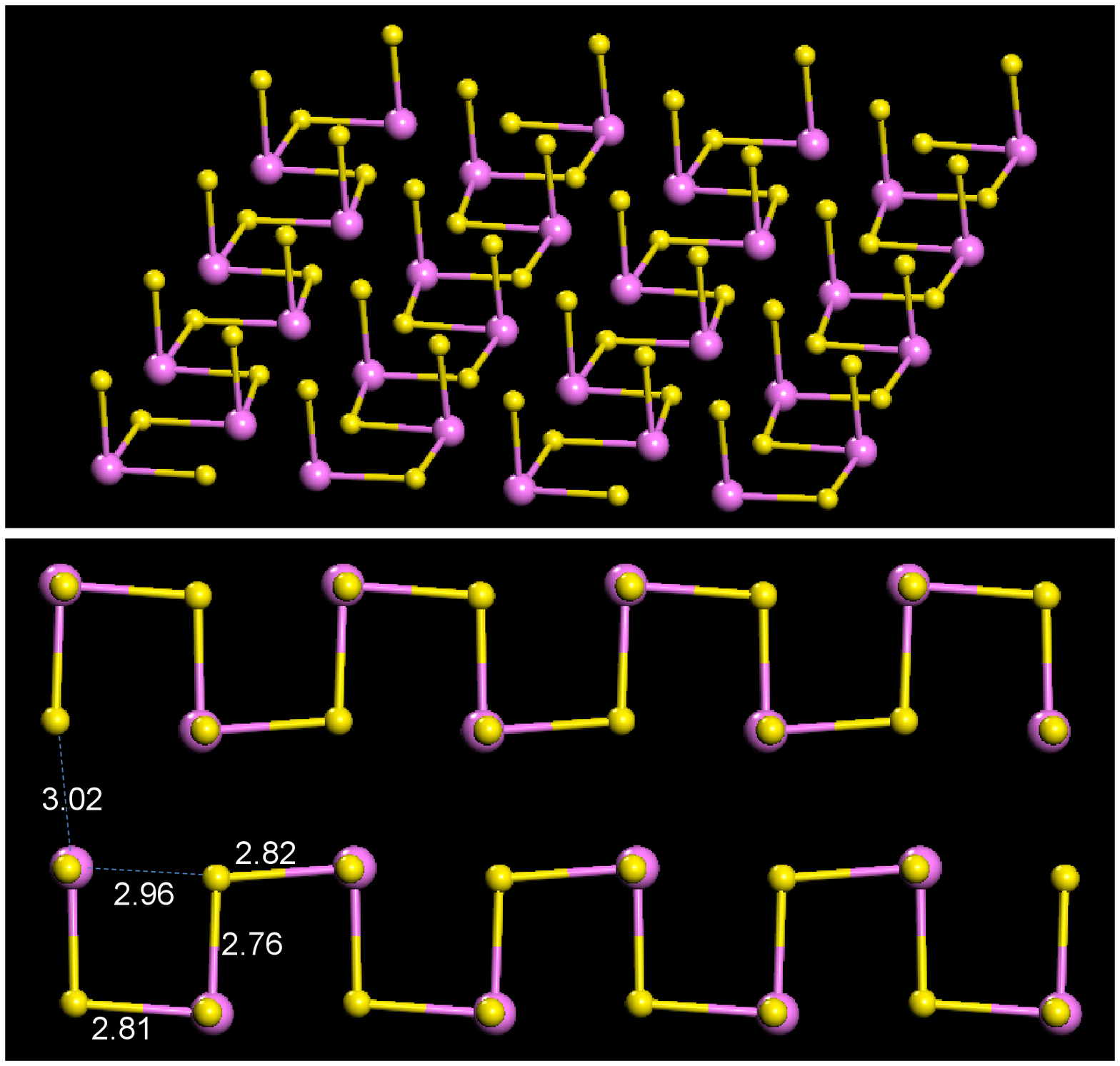} 
\includegraphics[height=5.0cm]{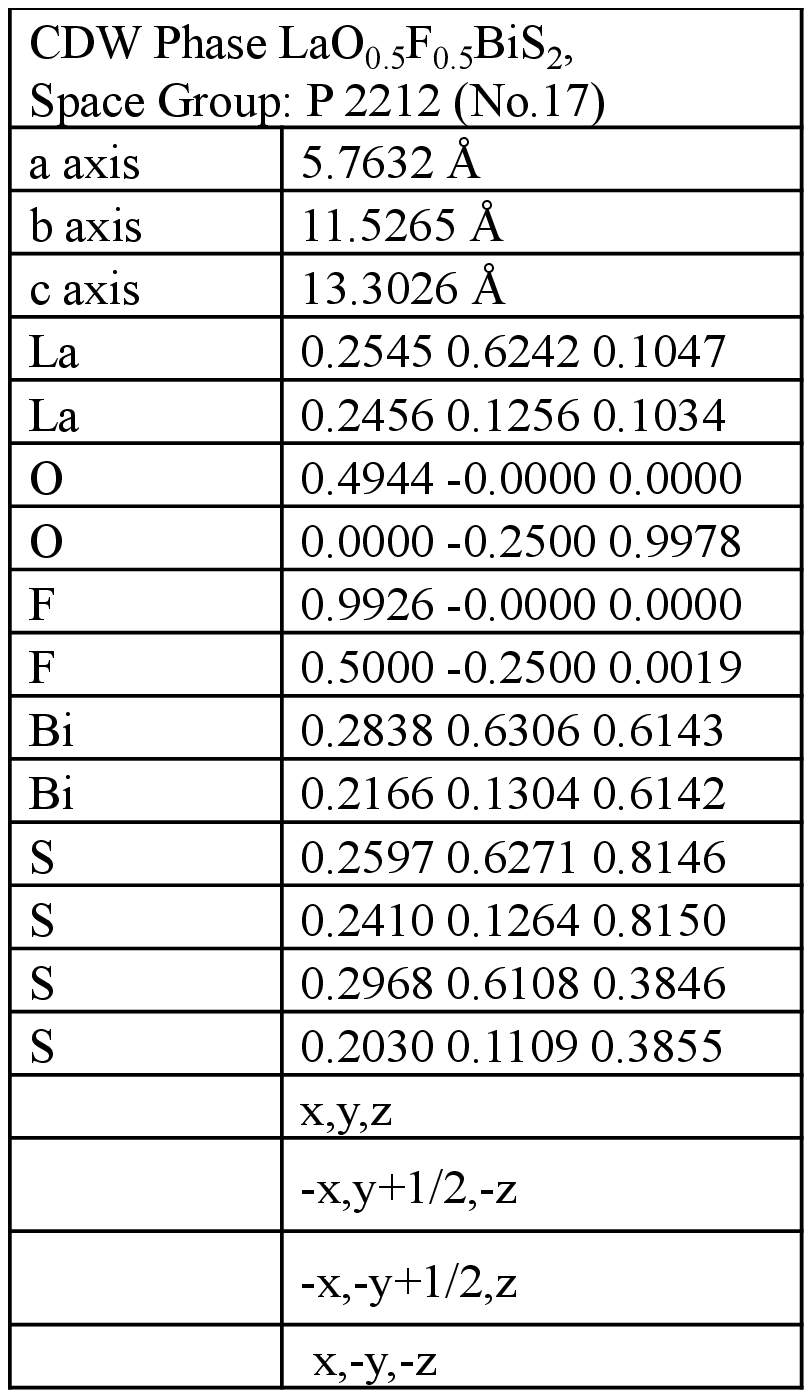}
\caption{
 The CDW phase of 
LaO$_{0.5}$F$_{0.5}$BiS$_2$ with the one-dimensional chains. 
Various Bi-S distances are also shown. On the right, 
we give the structural parameters and the symmetry 
operators of the CDW phase. 
}
\label{FigS4}
\end{figure}

\end{document}